\pdfoutput=1
\documentclass[12pt]{article}
\usepackage{epsfig, epsf, graphicx}
\usepackage{pstricks, pst-node, psfrag}
\usepackage{latexsym,amssymb,amsmath,bm,amsthm,amsfonts}
\usepackage{verbatim,enumerate}
\usepackage{rotating, lscape}
\usepackage{setspace}
\usepackage{multirow}
\usepackage{float}
\usepackage{wrapfig}
\usepackage{caption}
\usepackage{subcaption}
\usepackage[most]{tcolorbox}
\usepackage[round]{natbib}
\newtheorem{thm}{Theorem}
\newcommand{\bs}{\mathbf{s}}

\newcommand{\bu}{\mathbf{u}}
\DeclareMathOperator{\sign}{sign}
\begin{document}
\thispagestyle{empty} \baselineskip=28pt \vskip 5mm


\thispagestyle{empty} \baselineskip=28pt \vskip 5mm
\begin{center} {\Huge{\bf Vector Autoregressive Models with Spatially Structured Coefficients for Time Series on a Spatial Grid}}
\end{center}

\baselineskip=12pt \vskip 10mm

\begin{center}\large
	Yuan Yan\footnote[1]{
		\baselineskip=10pt 
		Department of Mathematics \& Statistics, Dalhousie University, Halifax, NS, B3H 4R2, Canada. \\
		E-mail: yuan.yan@dal.ca.},
	Hsin-Cheng Huang\footnote[2]{
		Institute of Statistical Science, Academia Sinica, Taipei 115, Taiwan. \\E-mail: hchuang@stat.sinica.edu.tw.}
	and Marc G.~Genton\footnote[3]{
		\baselineskip=10pt 
		Statistics Program, King Abdullah University of Science and Technology (KAUST), 
		Thuwal 23955-6900, Saudi Arabia. \\
		E-mail: marc.genton@kaust.edu.sa. \\ 
		This publication is based upon work supported by the King Abdullah University of Science and Technology
		(KAUST) Office of Sponsored Research (OSR) under Award No: OSR-2018-CRG7-3742 and Academia Sinica Investigator Award AS-IA-109-M05.}
\end{center}
\baselineskip=17pt \vskip 10mm \centerline{\today} \vskip 15mm

\begin{center}
	{\large{\bf Abstract}}
\end{center}
Motivated by the need to analyze readily available data collected in space and time, especially in environmental sciences, we propose a parsimonious spatiotemporal model for time series data on a spatial grid. 
In essence, our model is a vector autoregressive model that utilizes the spatial structure to achieve parsimony of autoregressive matrices at two levels. The first level ensures the sparsity of the autoregressive matrices using a lagged-neighborhood scheme. The second level performs a spatial clustering of the non-zero autoregressive coefficients such that within some subregions, nearby locations share the same autoregressive coefficients while across different subregions the coefficients may have distinct values. 
The model parameters are estimated using the penalized maximum likelihood with an adaptive fused Lasso penalty. The estimation procedure can be tailored to accommodate the need and prior knowledge of a modeler.
Performance of the proposed estimation algorithm is examined in a simulation study. 
Our method gives reliable estimation results that are interpretable and especially useful to identify geographical subregions, within each of which, the time series have similar dynamical behavior with homogeneous autoregressive coefficients. 
We apply our model to a wind speed time series dataset generated from a climate model over Saudi Arabia to illustrate its power in explaining the dynamics by the spatially structured coefficients. Moreover, the estimated model can be useful for building stochastic weather generators as an approximation of the computationally expensive climate model.  
\baselineskip=14pt
\par\vfill\noindent
{\bf Some key words:} Adaptive fused Lasso; Coefficients homogeneity; Penalized maximum likelihood; Regularization; Spatial clusters; Spatiotemporal model.
\par\medskip\noindent

\clearpage\pagebreak\newpage \pagenumbering{arabic}
\baselineskip=26pt
\clearpage\pagebreak\newpage \pagenumbering{arabic}
\baselineskip=26pt
\section{Introduction}
As spatiotemporal datasets are readily available in many scientific fields, various models have been proposed for modeling purposes. 
In this paper, we consider, in particular, time series data collected on a spatial grid, which are frequently encountered in the environmental, climate, and marine sciences. These data may be calculated from a satellite-borne instrument signal (e.g., scatterometer) or simulated from a physical model. Flexible and interpretable spatiotemporal models that are also parsimonious and computationally feasible are needed in order to understand the space-time dynamics underlying the data, make better forecasts, or generate fast simulation (stochastic weather generator) of such spatiotemporal data.

Traditional covariance-based space-time models are descriptive and are suitable for modeling data in continuous space and time, and allow interpolation (kriging) between locations and time points. However, these models consider no data generating mechanism. Alternatively,
\citet{DSTM} recommend dynamical spatiotemporal models (DSTM) over the descriptive spatiotemporal covariance modeling. One commonly adopted linear DSTM for describing temporal dynamics is the vector autoregressive (VAR) model, where each variable corresponds to the process at one spatial location. 
Here, we give a brief review of a VAR$(p)$ model in the DSTM framework. Throughout this paper, we assume $\bm{Z}_t=(Z_t(\bs_1),\dots,Z_t(\bs_n))'$ for $t=1,\dots,T$ is a time series of a zero-mean spatial process at $n$ locations. A Gaussian VAR$(p)$ model of dimension $n$ is defined as
\begin{equation}
\bm{Z}_t=\mathbf{A}_1\bm{Z}_{t-1}+\dots+\mathbf{A}_p\bm{Z}_{t-p}+\bm{\epsilon}_t, \quad t=2,\dots,T,
\label{eq:var1}
\end{equation}
where $\mathbf{A}_i$ are the $n \times n$ matrices of autoregressive coefficients (i.e., transition/propagator/redistribution matrices) of order $i$, and $\bm{\epsilon}_t=(\epsilon_t(\bs_1),\dots,\epsilon_t(\bs_n))'$ is a spatially correlated innovation process that follows an independent and identically distributed (i.i.d.) multivariate Gaussian distribution with mean $\mathbf{0}$ and covariance matrix $\mathbf{\Psi}$, $\bm{\epsilon}_t\overset{i.i.d.}{\sim}\mathcal{N}_n(\mathbf{0},\mathbf{\Psi})$, and $\mathbf{\Psi}$ could be completely unknown or modeled by some commonly used parametric spatial covariance function. 

The spatial setting for a VAR model can be both a curse and a blessing. 
On the one hand, for time series obtained at $n$ gridded locations, the number of parameters, $n^2$, for each transition matrix $\mathbf{A}_i$ can be enormous, which makes estimation difficult and unstable, if not impossible. 
On the other hand, by taking advantage of the spatial structure, parsimonious assumptions can be made for the transition matrices in order to get sensible estimation, and thus gain insights of the underlying dynamics of the phenomenon over space and time.

The main idea of our spatial VAR models are to enforce parsimony at two levels based on the gridded spatial structure.
First, the sparsity of each $\mathbf{A}_i$ is obtained by using a lagged-neighborhood scheme for a given set of $K_i$ neighbors. 
The non-zero coefficients in $\mathbf{A}_i$ can be interpreted as the influence or flow from a neighbor in a certain direction at time lag $i$.
Next, we assume the non-zero coefficients are homogeneous in space such that within some subregions, nearby locations share the same autoregressive coefficients while across different subregions the coefficients may have distinct values. 
For parameter estimation, the usual least-squares or maximum likelihood method suffer from high variances for time series data rich in space but not long in time, which ultimately lead to low interpretability. In order to impose coefficients homogeneity, we propose a penalized likelihood method for parameter estimation, with an adaptive fused Lasso penalty on the coefficients.

The originality of our model truly lies in the combination of sparsity, coefficients homogeneity, and the estimation procedure for flexible and interpretable spatial VAR models. 
Since coefficients homogeneity can be regarded as an extension of the concept of sparsity, the two-level specifications of our spatial VAR models achieve further dimension reduction for the coefficients than the sparse VAR models and therefore are capable of dealing with time series data rich in space (hundreds of locations, which is generally considered high-dimensional for a VAR model) but not long in time ($T \ll n^2$), and is especially useful to identify spatial subregions from the clustered coefficients. 

In Section~\ref{sec:lit}, we put the two-level parsimony assumptions of our spatial VAR models under the context of related and alternative modeling strategies in the literature.
In Section~\ref{sec:model}, we formalize our model for the VAR(1) case and develop as well as implement the parameter-estimation method. In Section~\ref{sec:sim}, we demonstrate the performance of our estimation algorithm in a simulation study. In Section~\ref{sec:app}, we apply our model to a daily wind speed dataset on a spatial grid covering Saudi Arabia generated from a climate model in order to illustrate its usefulness. Limitations and possible extensions of our model are discussed in Section~\ref{sec:diss}. Proofs and asymptotic properties of the penalized estimation can be found in the Appendix.

\section{Literature Review}\label{sec:lit}

For the first level, our spatial VAR models adopt a lagged-neighborhood scheme for a number of pre-specified neighbors (not necessarily nearest), which endows each $\mathbf{A}_i$ with some given sparsity structure.
The assumption of fixed sparsity structure for the transition matrices is simple while sensible and was adopted by multiple authors.
The hierarchical model in \citet{Wikle1998} included a VAR(1) process with a lagged-nearest-neighbor scheme for the transition matrix and was applied to the analysis of monthly averaged maximum-temperature data. 
\citet{Rao08} introduced a VAR($p$) model with a diagonal structure for each $\mathbf{A}_i$. 
\citet{Tagle17} adopted a VAR(2) process with a lagged-nearest-neighbor scheme for $\mathbf{A}_1$ and a diagonal structure for $\mathbf{A}_2$ to analyze wind speed data.

Rather than a fixed sparsity structure, a more flexible sparsity structure on the VAR transition matrices based on the data can be achieved by a variable selection/model building scheme or a regularized estimation method.
In econometrics, there is a vast literature on Bayesian VAR with different shrinkage priors on the vectorized transition matrices, e.g., Minnesota prior \citep{BVAR10}, adaptive hierarchical priors  \citep{BVAR19}, global-local shrinkage priors, i.e., Dirichlet-Laplace prior \citep{BVAR20}. 
However, these Bayesian VAR models only considered sparsity in the transition matrices and did not adapt to spatially referenced data.
For spatial data, \citet{LM05} proposed a variable selection procedure carried out at each location separately, in which the variables (locations) were ordered based on their relative distances (or importance, when additional information is available) from the location under study. First, they defined a spatial partial correlation function with respect to the order. Then they selected the model from the sample partial correlation functions, similarly to the order selection by checking the partial autocorrelation function (PACF) in the time series context. 
\citet{Hsu08} adopted the Lasso penalty for subset selection in a VAR model and estimated the non-zero coefficients simultaneously for a better forecasting performance. 
\citet{basu2015} provided a theoretical justification for estimating the VAR models with the Lasso penalty. 
\citet{Ngue15} used a weighted $L_1$ penalty as a function of the spatial distance between two locations for multilayer spatial VAR.
\citet{Add17} developed a two-step $L_1$ penalized least squares method to capture the sparsity structure of the transition matrices based on distance, while
\citet{MoAi17} applied the smoothly clipped absolute deviation (SCAD) penalty to facilitate a sparse structure on both the transition matrices and the precision matrix ($\mathbf{\Psi}^{-1}$) for the innovation process.

Other approaches to reduce the number of parameters involved in a spatial VAR include parametric modeling of the transition matrices and possibly combined with dimension reduction. 	
\citet{AiMo06} employed a VAR(1) model for wind fields with a latent process that described the motion of the air masses, where $\mathbf{A}_1$ and $\mathbf{\Psi}$ were both parameterized as a function of the latent translation of the wind fields. 
\citet{Wi01} and \citet{kat12} implemented a dimension reduction technique with spatial basis functions, and modeled the resulting lower-dimensional process using a VAR(1) model, where the transition matrix had a parametric block structure related to the spatial resolution of the spatial bases.  
\citet{bes15} built a spatiotemporal model in the state-space form, where the process at all points depended on a common univariate state variable, which was an AR$(1)$ time series that contained large-scale spatial variation (regional information). 

For the second level, our model assumes the non-zero coefficients that represent the influence from a neighbor in a certain direction are region-wise constant. Consequently, the dynamics of the process are the same within each subregions and are allowed to vary discontinuously across different subregions that are possibly with distinct geographical features. 
Previously, for linear regression, in order to identify homogeneous groups of predictors, \citet{GroupL10} used a nonconvex and overcomplete penalty on pairwise differences of the coefficients.
\citet{KeFan15} proposed a two-step procedure to first have preliminary ordered segments of the coefficients and then penalize the pairwise differences of the coefficients between adjacent segments and within individual segments.
When the predictors are naturally ordered, it becomes the fused Lasso method proposed by \citet{fused05}. The method penalizes the successive differences among the coefficients to achieve local constancy of the coefficients. \citet{adaLasso06} introduced the adaptive Lasso, by giving different weights adaptively for different penalty terms given the data. 
\citet{Sun16} used adaptive fused Lasso to estimate spatial and temporal quantile functions with spatially and temporally homogeneous groups, respectively.
Under the spatial regression setting, \citet{SpL10} proposed the spatial Lasso with homogeneous coefficients given known subregion structure. Recently, to identify the spatially clustered patterns in the coefficients, \citet{LiSang19} used the fused Lasso penalty on edges of the minimum spanning tree formed by the spatial locations.

Instead of the homogeneity approach, VAR models with smoothly varying diagonal coefficients were considered in \citet{Wikle1998} and \citet{Rao08}. 
\citet{Wikle1998} let the diagonal coefficients varied spatially according to a conditionally autoregressive (CAR) model for the spatial lattice while coefficients related to the four nearest neighbors are kept constant over space; 
while in \citet{Rao08}, the diagonal coefficients vary smoothly in space, which are estimated by localized least squares, in order to accommodate the spatial non-stationarity of the process and allow spatial interpolation. 
The above two models are the VAR extensions of the spatially varying coefficient processes \citep{SVC03, SVC18}, which themselves are spatial extensions of the varying coefficient regression models \citep{fan1999}.

Compared to the existing models, the novelty of our spatial VAR models is essentially to extend the pursuit of coefficients homogeneity to the spatial VAR setting, with the help of a pre-specified sparsity structure on the transition matrices that makes the non-zero coefficients physically meaningful with respect to the temporal dynamics.

\section{Methodology}\label{sec:model}
\subsection{The Spatial VAR(1) Model}
We formalize our model specifications for the simple VAR(1) case first to elucidate the essence of our model.
We let $D_s\subset\mathbb{R}^2$ be gridded locations, $D_s=\{\bs_1,\dots,\bs_n\}\subset\{(i_1,i_2):i_1=1,\dots,n_x,\,i_2=1,\dots,n_y\}$. 
We assume that for location $\bs_i$, $Z_t(\bs_i)$ is directly affected only by the lag-1 process at the $K$ neighboring locations (including the location itself), $Z_{t-1}(\bs_i+\bu_k), k=1,\dots,K$, for some translation vector $\bu_k$. For example, if $\bu_k=(-1,0)'$, then $\bs_i+\bu_k$ is the left neighbor of $\bs_i$, and we fix $\bu_1=\mathbf{0}$ so that the location $\bs_i$ itself is always included. 
For a finite grid, boundary conditions are required when $\bs_i+\bu_k\notin D_s$. We are only interested in coefficients and dynamics for the $n_I$ inner grid points, for each of which all $K$ neighbors are within $D_s$. 
Since the dynamics of the boundary locations are not of interest and are modelled only to assist estimation of the coefficients related to the inner grid points, a simple model can be assumed for the $n_B=n-n_I$ boundary points in order to simulate data on the boundary. 
For easy notation, we order the grid points such that the first $n_I$ points $\bs_i$ for $i=1,\dots,n_I$ are the inner grid points.  We adopt the lagged-neighborhood scheme only for these inner grid points; for the boundary grid points, we set $K=1$ so that their temporal evolution depends only on themselves, i.e. AR(1) model. 
This model can be expressed as	 	 
\begin{equation}
	\begin{aligned}
		Z_t(\bs_i)&=\sum_{k=1}^K \alpha_k(\bs_i)Z_{t-1}(\bs_i+\bu_k)+\epsilon_t(\bs_i),\quad &i=1,\dots,n_I,\,t=2,\dots,T, \\
		Z_t(\bs_i)&=\alpha_1(\bs_i)Z_{t-1}(\bs_i)+\epsilon_t(\bs_i),\quad &i=n_{I}+1,\dots,n,\,t=2,\dots,T,
	\label{eq:svar1}
	\end{aligned}
\end{equation}
where $\bm{\epsilon}_t\overset{i.i.d.}{\sim}\mathcal{N}_n(\mathbf{0},\mathbf{\Psi})$ and $\alpha_k(\bs_i)$ corresponds to the influence on the process at $\bs_i$ by its $k$th neighbor. 
The lagged-nearest-neighbor scheme in \citet{Wikle1998} and \citet{Tagle17} corresponds to the case when $K=5$, which takes account of the influence by the location itself and its four nearest neighbors. These five locations on a grid may be referred to as the five-point stencil in the numerical analysis literature, the first-order stencil neighborhood scheme in high-performance computing, and the rook neighbors or the first-order neighborhood elsewhere. It can be easily extended to the second-order neighborhood with $K=9$ (queen neighbors), which includes the nearest eight neighbors and the location itself. 
More flexibly, $\bu_k$ in $\eqref{eq:svar1}$ can be selected based on prior knowledge of the spatiotemporal process and the choices are not limited to the nearest neighbors. For example, if a river flows to the east, then it is sensible to model the chemical concentration at a location in the river to depend on only its western neighbors.

For each $k=1,\dots,K$, we let $\bm{\alpha}_k=(\alpha_k(\bs_1),\dots,\alpha_k(\bs_{n_I}))'$ be the vector of the unknown parameters of interest, which are the coefficients related to the $k$th neighbor for each of the $n_I$ inner grid points.
Let $\bm{\alpha}_\text{B}=(\alpha_1(\bs_{n_I+1}),\dots,\alpha_1(\bs_{n}))'$ be the vector of the unknown autoregressive coefficients of the boundary points.
Time series of a spatial process often have some regional structure such that locations within one region share the same spatiotemporal dynamics while the dynamics could vary discontinuously across different subregions.
Therefore, a spatial coefficient-grouping structure such that $\bm{\alpha}_k$ is composed of spatial clusters of values for each $k=1,\dots,K$ is desirable. We assume no spatial structure for $\bm{\alpha}_\text{B}$; again, these coefficients are of little practical importance and are only modeled to allow simulation and forecasting several steps ahead as well as to simplify notations.
Figure~\ref{fig:true} of our simulation design gives an intuitive visual illustration of the idea of spatial homogeneity in the autoregressive coefficients. 

The model defined by \eqref{eq:svar1} can be written as the VAR model in (\ref{eq:var1}) with order $p=1$ and $\mathbf{A}_1=\mathbf{A}$. 
We let $a_{ij}$ be the $(i,j)$th entry of $\mathbf{A}$. With the lagged-neighborhood scheme, $\mathbf{A}$ has a sparse form such that $a_{ij}\neq 0$, only if $\bs_j=\bs_i+\bu_k$ for some $k$ and there are at most a total of $m=K n_I + n_B$ non-zero elements in $\mathbf{A}$. 
Since the sparsity of $\mathbf{A}$ is combined with the spatial clustered structure of each $\bm{\alpha}_k$, the reduction in the number of parameters is drastic compared to the $n^2$ parameters required by the full VAR(1) model. 
We let $\bm{\alpha}=(\bm{\alpha}'_1,\dots,\bm{\alpha}'_K,\bm{\alpha}'_\text{B})'$ be the vector of all the unknown coefficients of length $m$; then, the sparse transition matrix $\mathbf{A}$ depends only on $\mathop{\bm{\alpha}}$ through $\mathrm{vec}(\mathbf{A})=\mathbf{P}\bm{\alpha}$, where $\mathbf{P}$ is a $n^2 \times m$ matrix of zeros and ones that indicates the positions of the corresponding non-zero elements in $\mathrm{vec}(\mathbf{A})$ for $\bm{\alpha}$. 



\subsection{Parameter Estimation by the Adaptive Fused Lasso}\label{sec:est}	
For our model, the dynamical spatiotemporal coefficients $\bm{\alpha}$ and the matrix $\mathbf{\Psi}$ that models the spatial covariance of the residuals need to be estimated. To capture the spatially homogeneous groups of $\bm{\alpha}$, we consider a penalized likelihood method with the adaptive fused Lasso penalty.
\citet{genlasso11} developed a path algorithm for a `generalized' Lasso problem that included the Lasso, fused Lasso, and adaptive Lasso as special cases. This generalized Lasso problem can be solved efficiently and is easy to implement using the `genlasso' package \citep{genlassoR} in R \citep{R19}.


We consider the log-likelihood of $\bm{Z}_2,\dots,\bm{Z}_T$ conditioning on $\bm{Z}_1$ for our model with $\mathbf{\Psi}$ given. For any matrix $\bm{X}$ and vector $\bm{y}$ that satisfy
\begin{align}
{\bm{X}'\bm{X}}
=&~ \mathbf{P}'\left\{\left(\sum_{t=2}^{T}\bm{Z}_{t-1}\bm{Z}_{t-1}'\right)\otimes\mathbf{\Psi}^{-1}\right\}\mathbf{P},\label{eq:x}\\
{\bm{X}'\bm{y}}
=&~ \mathbf{P}'\mathrm{vec}\left(\mathbf{\Psi}^{-1}\sum_{t=2}^{T}\bm{Z}_{t}\bm{Z}_{t-1}'\right),\label{eq:y}
\end{align} 
the conditional log-likelihood can be written as a second-order polynomial of $\bm{\alpha}$ (the proof can be found in the Appendix): 
\begin{align}
l(\bm{\alpha})=-\frac{1}{2}\sum_{t=2}^T(\bm{Z}_t-\mathbf{A}\bm{Z}_{t-1})'\mathbf{\Psi}^{-1}(\bm{Z}_t-\mathbf{A}\bm{Z}_{t-1})
=-\frac{1}{2}\big\|\bm{y}-\bm{X}\bm{\alpha}\big\|_2^2+\mathrm{constant},
\label{eq:form}
\end{align}
where $\|\cdot\|_2$ is the $\ell_2$-norm. 
One way to choose $\bm{X}$ is to compute the Cholesky decomposition of the matrix on the right-hand side of (\ref{eq:x}) and, then, solve for $\bm{y}$ from (\ref{eq:y}).  

We consider a penalized maximum likelihood (PML) method with the following objective function:
\begin{align}
F(\bm{\alpha})
=&~ l(\bm{\alpha}) -\lambda\sum_{k=1}^K\sum_{{i}\sim {j}}w_{k,{i},{j}}|\alpha_k(\bs_{i})-\alpha_k(\bs_{j})|,
\label{eq:Phi}
\end{align}
where the last term is an adaptive 2D fusion problem penalty that facilitates the grouping of neighboring coefficients,
${i}\sim {j}$ denotes that $\bs_{i}$ is a direct neighbor of $\bs_{j}$ (there is an edge between $\bs_{i}$ and $\bs_{j}$ on the grid), for $i,j= 1,\dots,n_I$,  $\lambda$ is a tuning parameter that controls the degree of grouping, and $\{w_{k,{i},{j}}\}$ are known weight values set to $w_{k,{i},{j}}=|\tilde{\alpha}_k(\bs_i)-\tilde{\alpha}_k(\bs_j)|^{-\gamma}$, where $\tilde{\bm{\alpha}}$ is a root-$T$ consistent estimator of $\bm{\alpha}$ and $\gamma$ is some fixed positive constant. 
We penalize the pairwise difference of $\alpha_k(\cdot)$ between direct neighbors within the inner grid points. For each $k=1,\ldots,K$, the number of penalty terms equals to the number of edges in the inner grid. Instead of the extensive pairwise difference penalty used by \citet{GroupL10} or the difference between the coefficient of a location and the average of its neighboring coefficients used by \citet{Sun16}, we take advantage of the spatial configuration and penalize on `just enough' pairs for coefficients clustering. Even though additional penalty terms could be put on individual coefficients to encourage some coefficients to be exactly zeros, with penalty terms only on the pairwise differences estimation of true zero coefficients can be close enough to zero and avoid extra computational difficulty. 

The maximization problem (\ref{eq:Phi}), given $\mathbf{\Psi}$, is a generalized Lasso problem, 
\begin{equation}
\min_{\bm{\alpha}}\left\{ \frac{1}{2}\left\lVert \bm{y}-\bm{X}\bm{\alpha} \right\rVert_2^2+\lambda\left\lVert \mathbf{D}\bm{\alpha} \right\rVert_1\right\}, \label{eq:lasso}
\end{equation}
where $\|\cdot\|_1$ is the $\ell_1$-norm for some penalty matrix $\mathbf{D}$, which is determined by the grid structure as well as $w_{k,i,{j}}$. 
The neighborhood structure of $\bm{\alpha}$ given by \eqref{eq:Phi} and \eqref{eq:lasso} corresponds to a graph formed by $K$ disjoint subgraphs with the same grid structure;
$K$ determines the level of sparsity for $\mathbf{A}$, while the penalty on pairwise difference between direct neighbors assists in capturing the spatial homogeneity of the VAR coefficients. 

When $\lambda=0$, the solution of \eqref{eq:lasso} corresponds to the restricted generalized least squares (GLS) estimator of $\mathbf{A}$ with the constraints that $a_{ij}=0$ except for the $m$ parameters corresponding to $\bm{\alpha}$.
As $\lambda\rightarrow\infty$, $\alpha_k(\cdot)$ reduces to a constant function because the network corresponding to $\{\bs_i: i=1,\dots, n_I\}$ is connected.
A proper tuning parameter $\lambda$ can be selected by 5-fold cross validation \citep{Ber12} or by the Bayesian information criterion (BIC), BIC$(\lambda)= \left\lVert \bm{y}-\bm{X}\bm{\alpha}_\lambda \right\rVert_2^2 + (T-1)\log(|\mathbf{\Psi}_\lambda|) + \log(T-1)\left\{\text{df}(\bm{X}\bm{\alpha}_\lambda)+\text{df}(\mathbf{\Psi})\right\}$, where $\bm{\alpha}_\lambda$ is the solution to the generalized Lasso problem in \eqref{eq:lasso} for a given $\lambda$, $\mathbf{\Psi}_\lambda$ is an estimator of ${\mathbf{\Psi}}$ given $\bm{\alpha}_\lambda$, 
$\text{df}(\bm{X}\bm{\alpha}_\lambda)$ can be estimated from the number of distinct values in $\bm{\alpha}_\lambda$ \citep{genlasso11}, and $\text{df}(\mathbf{\Psi})$ is the number of parameters in estimating $\mathbf{\Psi}$.
We can estimate $\mathbf{\Psi}$, given $\bm{\alpha}$ and the corresponding $\mathbf{A}$, using the maximum likelihood estimator:
\begin{equation}
\hat{\mathbf{\Psi}}=\frac{1}{T-1}\sum_{t=2}^T(\bm{Z}_t-\mathbf{A}\bm{Z}_{t-1})(\bm{Z}_t-\mathbf{A}\bm{Z}_{t-1})'.
\label{eq:psi}
\end{equation}
However, there is a trade-off in estimating $\mathbf{A}$ and $\mathbf{\Psi}$, namely they compete to account for the spatial correlation. When $\mathbf{\Psi}$ is difficult to estimate, it could be assumed to be diagonal and estimated accordingly. 
For very short time series when $T<n$, the non-parametric estimation of $\mathbf{\Psi}$ in \eqref{eq:psi} is singular. In this case, further regularization can be imposed on $\mathbf{\Psi}$ or a parametric model can be adopted to estimate $\mathbf{\Psi}$, such as the Mat\'ern covariance family. 




We propose the following algorithm for estimating $\bm{\alpha}$ and $\mathbf{\Psi}$ iteratively in two steps to get the adaptive fused Lasso estimator. We start with $\mathbf{\Psi}=\mathbf{I}_n$ and solve the PML \eqref{eq:lasso} with uniform weights and $\lambda$ selected by the BIC to get the fused Lasso estimator. The restricted OLS estimator could be obtained by setting $\lambda=0$ in step I. Then in step II, we compute weights from the fused Lasso estimator and solve the PML \eqref{eq:lasso} with $\mathbf{\Psi}$ estimated from the residuals $\bm{Z}_t-\hat{\mathbf{A}}^{(1)}\bm{Z}_{t-1}$ by \eqref{eq:psi} to obtain the adaptive fused Lasso estimator.

	\begin{tcolorbox}[breakable, enhanced]
		\textsc{Algorithm:}
		\begin{enumerate}
			\item Step I: Fused Lasso and restricted OLS
			\begin{enumerate}
				\item Fix $\mathbf{\Psi}=\hat{\mathbf{\Psi}}^{(0)}=\mathbf{I}_n$ and set $w_{k,i,j}\equiv 1$, solve \eqref{eq:lasso} with a sequence of $\lambda$ to obtain the fused Lasso estimators $\hat{\bm{\alpha}}^{(1)}_\lambda$ (where $\lambda=0$ corresponds to the restricted OLS estimator with the constraints that $a_{ij}=0$ except for the $m$ parameters corresponding to $\bm{\alpha}$); 
				\item For each $\lambda$, find and fix the corresponding $\mathbf{A}=\hat{\mathbf{A}}_\lambda^{(1)}$ and get $\hat{\mathbf{\Psi}}_\lambda^{(1)}$ from \eqref{eq:psi}, then calculate BIC with $\hat{\bm{\alpha}}_\lambda^{(1)}$ and $\hat{\mathbf{\Psi}}_\lambda^{(1)}$;	
				\item Select $\lambda$ by BIC and get the fused Lasso estimation $\hat{\bm{\alpha}}^{(1)}$, the corresponding $\hat{\mathbf{A}}^{(1)}$ and $\hat{\mathbf{\Psi}}^{(1)}$;
			\end{enumerate}
			\item Step II: Adaptive fused Lasso and restricted GLS
			\begin{enumerate}
				\item Fix $\mathbf{\Psi}=\hat{\mathbf{\Psi}}^{(1)}$ and update $w_{k,i,j}=|\hat{\alpha}^{(1)}_k(\bs_i)-\hat{\alpha}^{(1)}_k(\bs_j)|^{-1}$, to obtain the adaptive fused Lasso estimators $\hat{\bm{\alpha}}_\lambda$ by solving~(\ref{eq:lasso}) with a sequence of $\lambda$ (where $\lambda=0$ corresponds to the restricted GLS estimator);
				\item For each $\lambda$, find and fix the corresponding $\mathbf{A}=\hat{\mathbf{A}}_\lambda$ and get $\hat{\mathbf{\Psi}}_\lambda$ from \eqref{eq:psi}, then calculate BIC with $\hat{\bm{\alpha}}_\lambda$ and $\hat{\mathbf{\Psi}}_\lambda$;	
				\item Select $\lambda$ by BIC and get the adaptive fused Lasso estimation $\hat{\bm{\alpha}}$, the corresponding $\hat{\mathbf{A}}$ and $\hat{\mathbf{\Psi}}$.
			\end{enumerate}
		\end{enumerate}
	\end{tcolorbox}	


This adaptive fused Lasso estimator possesses the asymptotic properties of consistency in grouping and asymptotic normality. Details and proof are provided in the Appendix.

\subsection{Extensions}
Extending to the general VAR($p$) case, our model assumes for $j=1,\dots,p$, the process at each of the $i=1,\dots,n_I$ inner grid points is directly affected only by the lag-$j$ process at the prescribed $K_j$ neighboring locations. 
For boundary grid points, an AR($p$) process is assumed. The model can be written as
\begin{equation*}
	\begin{aligned}
		Z_t(\bs_i)&=\sum_{j=1}^{p}\sum_{k=1}^{K_j} \alpha_{jk}(\bs_i)Z_{t-j}(\bs_i+\bu_{k})+\epsilon_t(\bs_i), \quad &i=1,\dots,n_I,\,t=p+1,\dots,T, \\
		Z_t(\bs_i)&=\sum_{j=1}^{p}\alpha_j(\bs_i)Z_{t-j}(\bs_i)+\epsilon_t(\bs_i),\quad &i=n_{I}+1,\dots,n,\,t=p+1,\dots,T.
	\end{aligned}
\end{equation*}
Again, the lagged-neighborhood scheme ensures sparsity of each of the transition matrices.
Then the homogeneity assumption is on $\bm{\alpha}_{jk}=(\alpha_{jk}(\bs_1),\dots,\alpha_{jk}(\bs_{n_I}))'$ for each $j=1,\dots,p$ and $k=1,\dots,K_j$. 
Detailed equations and estimation algorithm can be derived similarly as the VAR(1) case.

Moreover, instead of using a fixed order $p$ and pre-defined neighboring scheme, order and neighbor selections can be made simultaneously using an additional Lasso step or the various variable selection methods mentioned in Section \ref{sec:lit}.



\section{Simulation Study}\label{sec:sim}
\subsection{Simulation Design}
\begin{figure}[b!] 
	\centering
	\includegraphics[width=\textwidth]{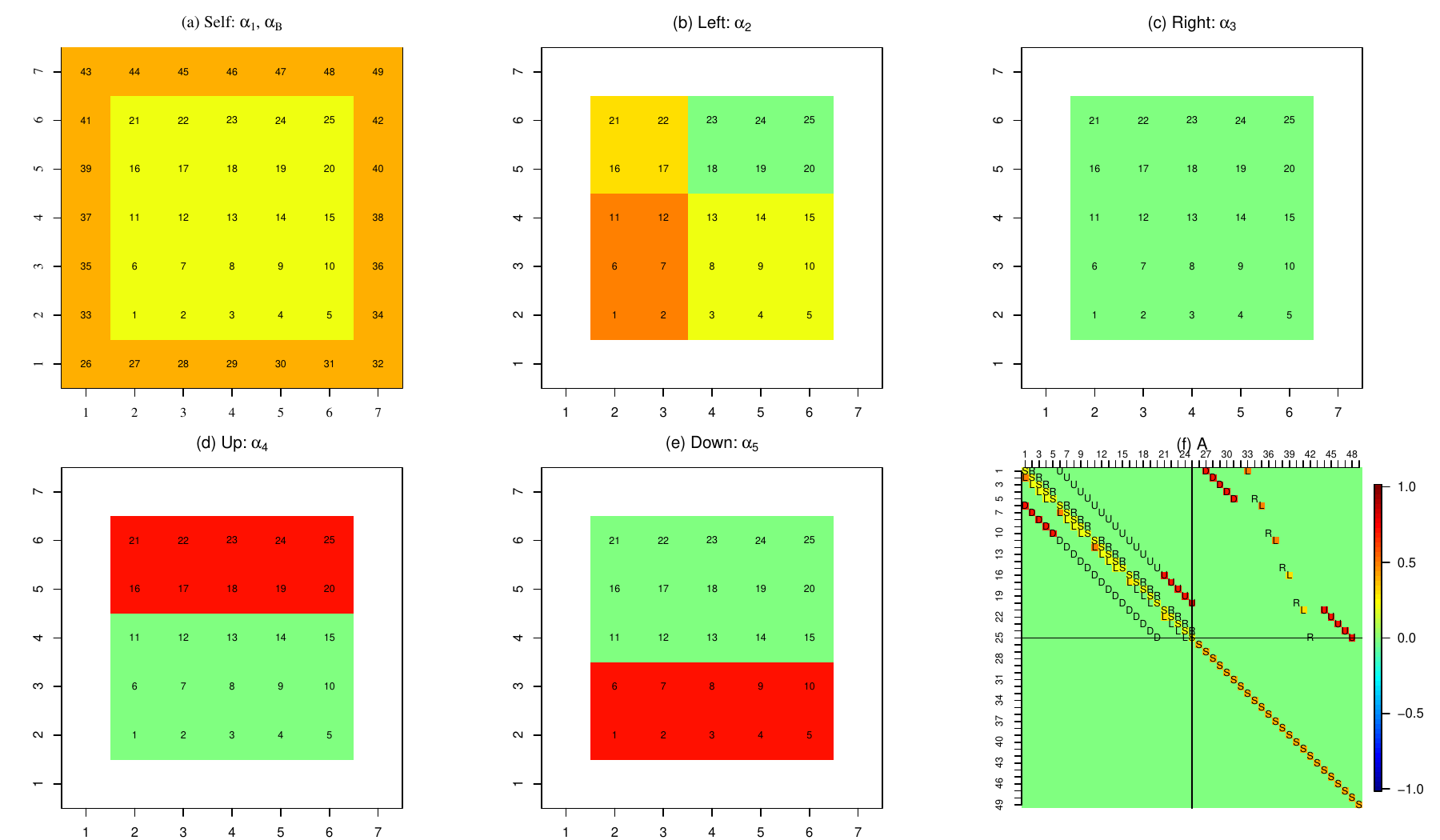}	
	\caption{(a) Values of $\bm{\alpha}_1$ (yellow) and $\bm{\alpha}_\text{B}$ (orange), and the indices of the 7 $\times$ 7 grid; (b) values of $\bm{\alpha}_2$; (c) values of $\bm{\alpha}_3$; (d) values of $\bm{\alpha}_4$; (e) values of $\bm{\alpha}_5$; (f) values of  $\mathbf{A}$ with its elements corresponding to the indices in (a).}
	\label{fig:true}
\end{figure}

We examine the performance of our estimation algorithm in a simulation study. We simulate time series data on a 7 $\times$ 7 grid ($n=49$) in a unit square from our model \eqref{eq:svar1} with $T\in\{100,500\}$ and $K=5$, where $\bm{\alpha}=(\bm{\alpha}'_1,\dots,\bm{\alpha}'_5,\bm{\alpha}'_{\text{B}})'$ and the corresponding $\mathbf{A}$ are shown in Figure~\ref{fig:true}. The spatially correlated error term $\bm{\epsilon}_t$ is generated from a multivariate Gaussian distribution with mean $\mathbf{0}$, and the covariance matrix $\mathbf{\Psi}$ is determined by an exponential covariance function with unit variance and a range parameter of 0.25. 
For this 7 $\times$ 7 square grid, we have $n_I=25, n_B=24$, and thus, $m=149$ for the length of $\bm{\alpha}$. 
We design the structure of $\bm{\alpha}$ such that $\{\bm{\alpha}'_1,\dots,\bm{\alpha}'_5\}$ have different numbers of clusters (i.e., distinct elements). For example, $\bm{\alpha}_4$ and $\bm{\alpha}_5$, corresponding to the neighbors above and below, have two clusters which mimic effects by a mountainous region in the middle separating the two subregions above and below; $\bm{\alpha}_2$ corresponding to the left neighbor has four clusters; $\bm{\alpha}_1$ corresponding to the location itself is a constant vector; and $\bm{\alpha}_3$ corresponding to the right neighbor is constant $\bm{0}$. 

We estimate $\mathbf{A}$ and $\mathbf{\Psi}$ using the proposed algorithm to get the adaptive fused Lasso estimator. For comparison, we also consider the restricted OLS estimator (with $\lambda=0$ in step I(a)), the restricted  GLS estimator (with $\lambda=0$ in step II(a)), and the fused Lasso estimator $\hat{\bm{\alpha}}^{(1)}$ (in step I(c)) of our algorithm.

\subsection{Estimation Results}	
\begin{figure}[t!] 
\centering
\includegraphics[width=\textwidth]{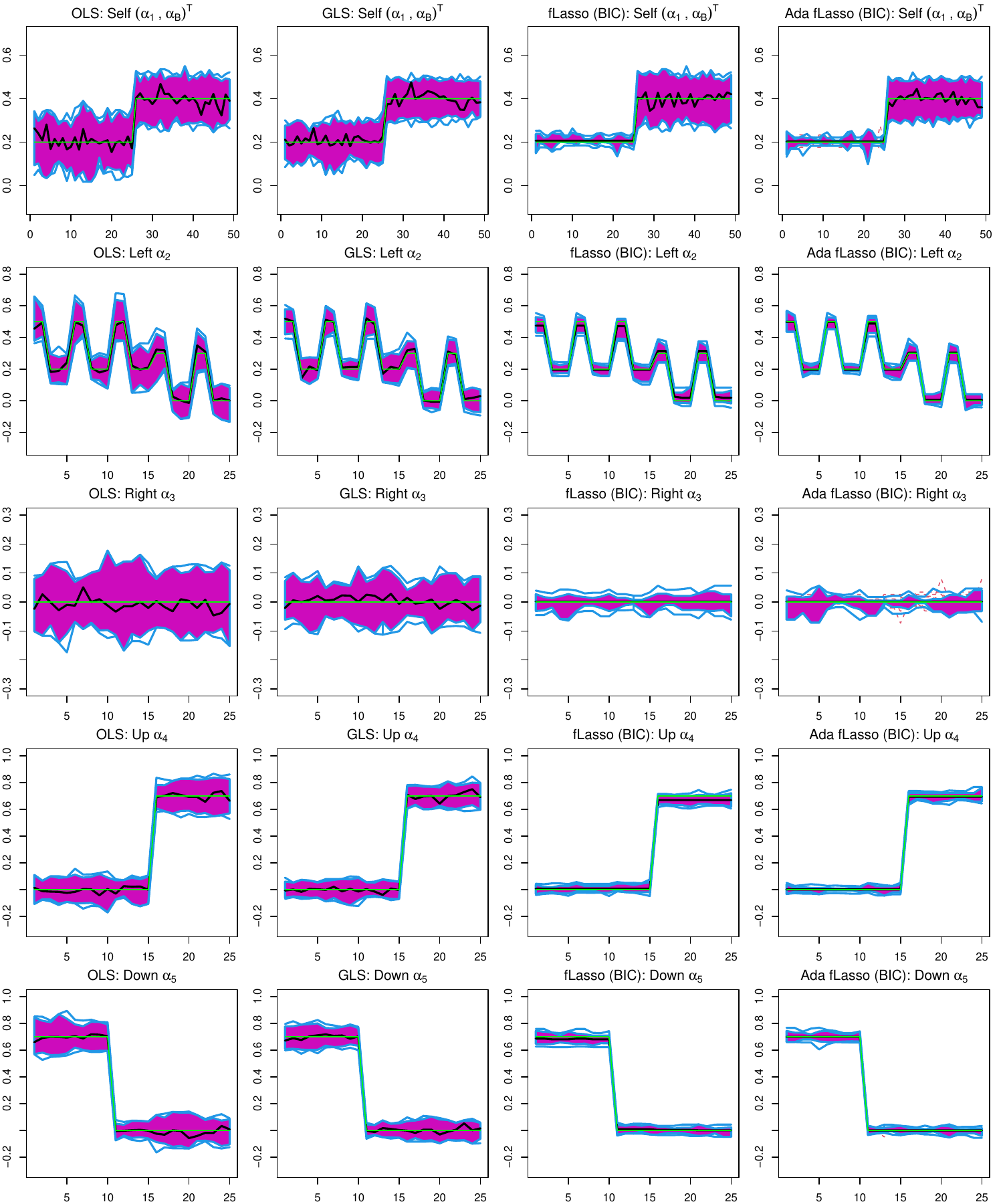}	
\caption{Functional boxplots of the estimated $\bm{\alpha}_k, k=1,\dots,5$ using restricted OLS, restricted  GLS, fused Lasso, and adaptive fused Lasso methods with $K=5$ and $T=500$ from 500 simulations. True values for $\bm{\alpha}_k$ are indicated by the green lines, where the order of the coefficients $\bm{\alpha}_k$ is indicated in the corresponding panel of Figure~\ref{fig:true}.}
\label{fig:est}
\end{figure}

Even though the structure of the coefficients is best illustrated two-dimensionally as in Figure~\ref{fig:true}, we can `stretch' (vectorize) them into a single vector ordered by the numbers labeled in each panel of Figure~\ref{fig:true}. Then, we could use the functional boxplot \citep{SG11} to visualize the performance of the estimators based on 500 simulation replicates. Figure~\ref{fig:est} shows the functional boxplots for the estimated coefficients of the five neighbors (in five rows) by the four methods (in four columns). 

From Figure~\ref{fig:est}, we see that both the fused Lasso and the adaptive fused Lasso estimators have much smaller variances and are, therefore, more stable than the restricted OLS and GLS estimators. 
The functional boxplot is designed to capture the structure of the estimated curve as a whole instead of estimation at individual points. Comparing the median curves of each plot, we see that the non-regularized methods do not reproduce the structure of the coefficients, while the fused Lasso and the adaptive fused Lasso both capture the grouping of the coefficients very well. There is a small departure of the median curves from the green lines for the fused Lasso, and the adaptive fused Lasso fixes this bias problem. 

Although ${\bm{\alpha}}_3=\bm{0}$, the estimates of ${\bm{\alpha}}_3$ from our adaptive fused Lasso are only very close to but not exactly zeros.
\citet{fused05} used a sparse fused Lasso that penalized both the coefficients and the pairwise differences of the coefficients to enforce both sparsity and grouping. However, in our algorithm, we only penalize the pairwise differences, not the coefficients themselves, i.e., the 2D fusion problem. In an extra simulation (results not shown), we used the sparse fused Lasso estimator, which can also be written in the form of \eqref{eq:lasso}, where $\mathbf{D}$ has $K n_I$ additional rows added to penalize the coefficients themselves. We found that our adaptive fused Lasso method without the Lasso sparsity penalization not only performs as satisfactorily as the sparse fused Lasso method, but it is also much more efficient in computation.

\begin{figure}[t!] 
\centering
\includegraphics[width=\textwidth]{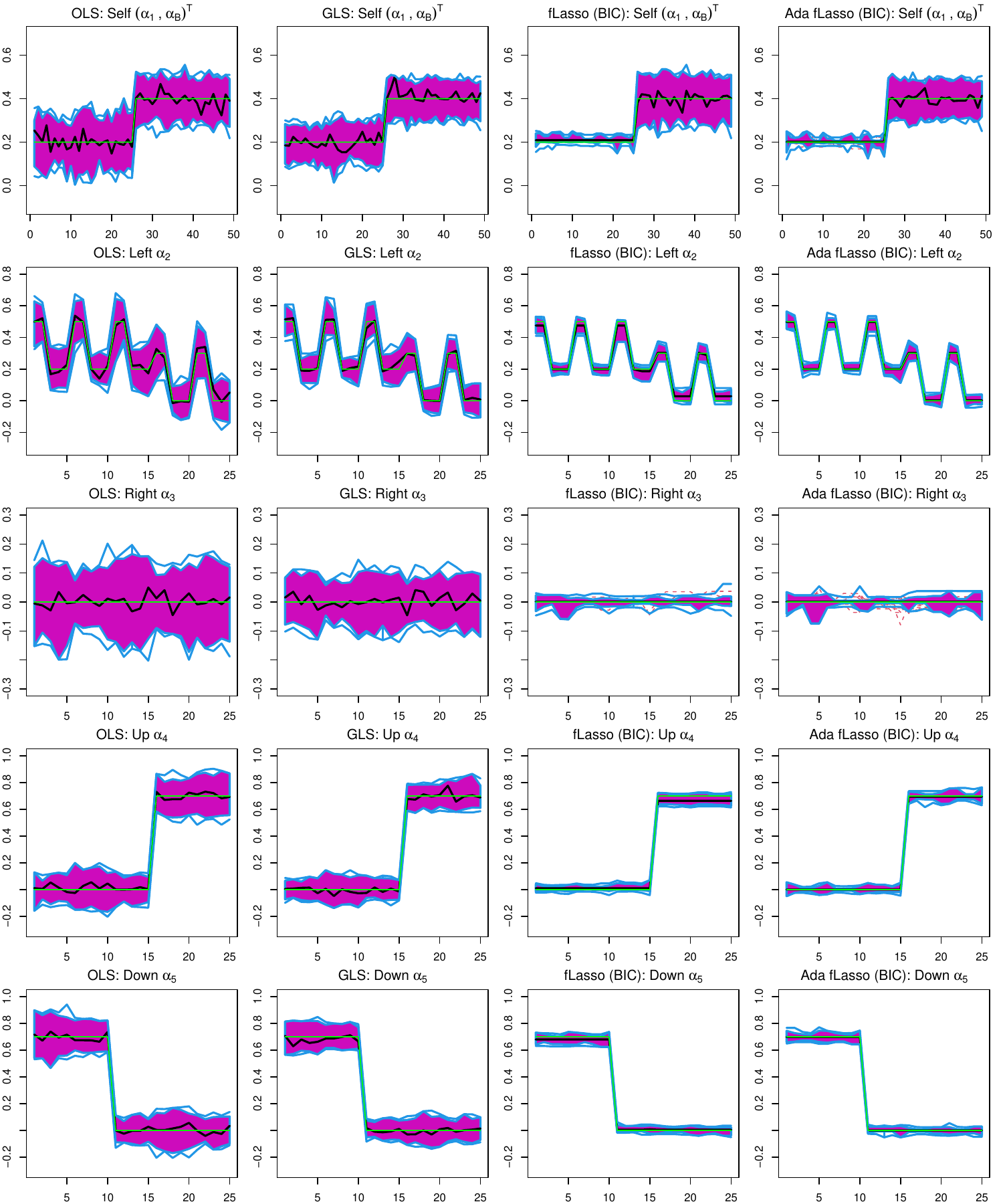}
\caption{Functional boxplots of the estimated $\bm{\alpha}_k, k=1,\dots,5$ using restricted OLS, restricted  GLS, fused Lasso, and adaptive fused Lasso methods with $K=9$ and $T=500$ from 500 simulations. True values for $\bm{\alpha}_k$ are indicated by the green lines, where the order of the coefficients $\bm{\alpha}_k$ is indicated in the corresponding panel of Figure~\ref{fig:true}.}
\label{fig:est8}
\end{figure}

\begin{figure}[t!] 
\centering
\includegraphics[width=\textwidth]{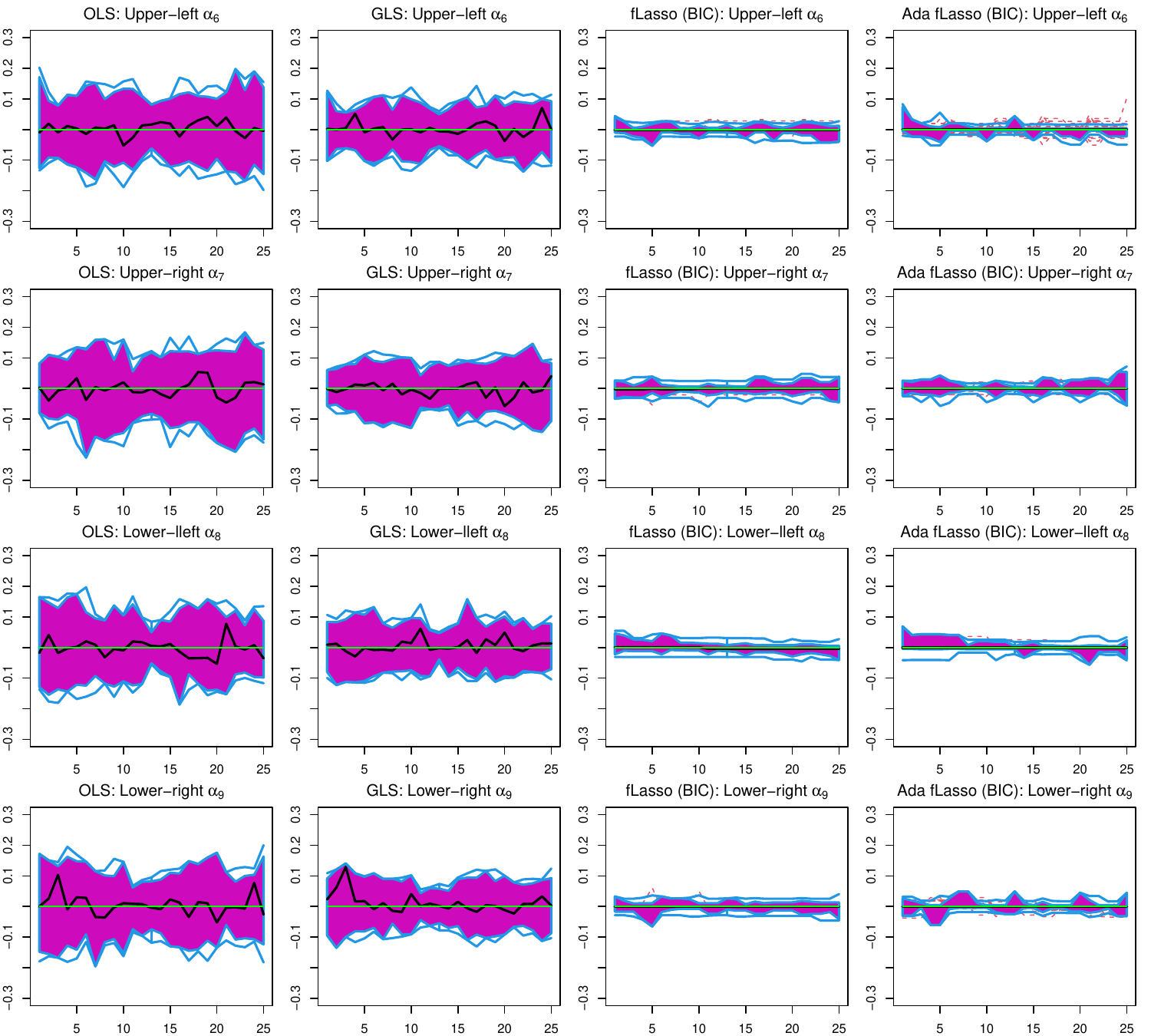}
\caption{Functional boxplots of the estimated $\bm{\alpha}_k, k=6,\dots,9$ using restricted OLS, restricted  GLS, fused Lasso, and adaptive fused Lasso methods with $K=9$ and $T=500$ from 500 simulations. True values for $\bm{\alpha}_k$ are indicated by the green lines, which are all zeros.}
\label{fig:est82}
\end{figure}

\begin{figure}[b!] 
\centering
\includegraphics[width=0.8\textwidth]{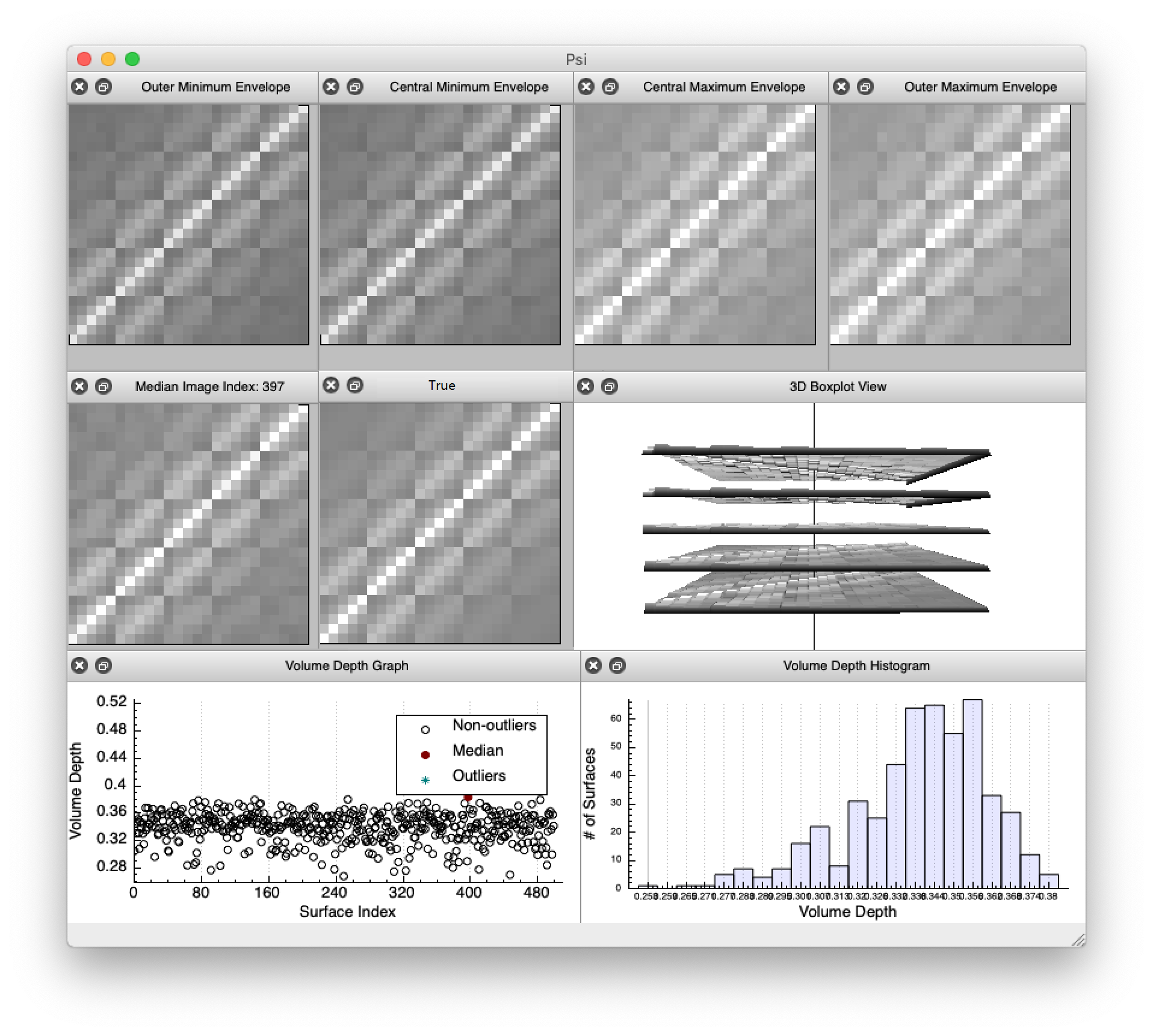}	
\caption{Surface boxplot of the estimated  $\mathbf{\Psi}$ using our algorithm with $T=500$ from 500 simulations.}
\label{fig:est_psi}
\end{figure}

Though the choice of $K$ might seem subjective, we can always specify a larger $K$ until the residuals pass the standard model validation procedure for a VAR model. Using a larger $K$ increases the computational effort, but does not significantly affect the performance of the non-zero coefficients since the adaptive fused Lasso already enforces sparsity in the coefficients. Figure~\ref{fig:est8} shows estimation results with $K=9$ in the same simulation design as described before. Comparing Figure~\ref{fig:est8} to Figure~\ref{fig:est}, we see that the estimates of $\bm{\alpha}_k,k=1,\dots,5$, by the fused Lasso and the adaptive fused Lasso with $K=9$ are similar to those with $K=5$, whereas the variances of the unregularized estimators by the restricted OLS and the restricted GLS are slightly expanded. Functional boxplots of the estimated $\bm{\alpha}_k$ (where the true $\bm{\alpha}_k$ is $\bm{0}$) for $k=6,\dots,9$, are shown in Figure~\ref{fig:est82}. Again, we see the restricted OLS estimates and the restricted GLS estimates have high variances, whereas both the fused Lasso and the adaptive fused Lasso can capture the sparsity in $\bm{\alpha}$ much better.

The pattern of the estimated $\mathbf{\Psi}$ is intrinsically two-dimensional; therefore, we resort to the surface boxplot \citep{sufbplot}, which is the two-dimensional version of the functional boxplot, to demonstrate the estimation results. Figure~\ref{fig:est_psi} shows the surface boxplot for the proposed estimator of $\mathbf{\Psi}$ based on 500 simulation replicates. We see that the median image is very close to the true structure, even though our method is somewhat non-parametric.

This simulation study shows that the structure of the coefficients in our model can be captured and estimated quite well by the proposed estimation algorithm, despite the fairly small number of time points $T$ relative to the number of parameters $n^2+n(n+1)/2$ to be estimated for a conventional VAR(1) model with dimension $n$.

\subsection{Forecast Results}
\begin{table}[b!]
\centering
\caption{PMSE of $h$-step ahead forecasts for $h=1,2,3$ based on four estimates of $\mathbf{A}$ from 500 simulations for both the 49 locations and the inner 25 points; the standard errors are given in parentheses.}	
\begin{tabular}{c|c|c|c|c|c}
	& $h$ & Restricted OLS & Restricted GLS & Fused Lasso & Adaptive fused Lasso\\ \hline\hline
	\multirow{6}{*}{49 locations}
	& \multirow{2}{*}{1} & 1.0060 & 1.0033 & 1.0025&  1.0012 \\
	& & (0.00012) & (0.00006) & (0.00006) & (0.00003) \\\cline{2-6}
	& \multirow{2}{*}{2}  & 1.0035 & 1.0020 & 1.0018 &  1.0008\\
	& & (0.00008) & (0.00004) & (0.00004) &  (0.00002) \\\cline{2-6}
	& \multirow{2}{*}{3}  & 1.0019 & 1.0011 & 1.0010 &  1.0005 \\
	& &  (0.00005) & (0.00003) & (0.00003) &  (0.00001) \\\hline
	\multirow{6}{*}{25 inner}
	&\multirow{2}{*}{1} & 1.0098 & 1.0052 & 1.0030 &  1.0010\\
	& & (0.00021) & (0.00010) & (0.00008) & (0.00004) \\\cline{2-6}
	& \multirow{2}{*}{2}& 1.0058 & 1.0032 & 1.0025 &  1.0010 \\
	points & & (0.00013) & (0.00007) & (0.00007) & (0.00002) \\\cline{2-6}
	&\multirow{2}{*}{3}& 1.0033 & 1.0018 & 1.0017 &  1.0007 \\
	& &(0.00009) & (0.00005) & (0.00005) & (0.00002) \\\hline
\end{tabular}
\label{forecast}
\end{table}

We compare the $h$-step-ahead forecasts of a VAR(1) model for $h=1,2,3$ at all grid points when using the estimated $\mathbf{A}$ by the four methods under consideration (i.e., restricted OLS, restricted  GLS, fused Lasso, and adaptive fused Lasso). To measure the forecasting performance, we consider the normalized $h$-step prediction mean squared error (PMSE), defined as \citep[modified from][]{Hsu08}:
\begin{align}
\text{PMSE}&=\frac{1}{n}\mathbb{E}\left[(\bm{Z}_{t+h}-\hat{\bm{Z}}_{t+h})'\bm{\Sigma}_h^{-1}(\bm{Z}_{t+h}-\hat{\bm{Z}}_{t+h})\right] \nonumber \\ 
&=1+\frac{1}{49}\mathbb{E}\left[(\hat{\bm{Z}}^*_{t+h}-\hat{\bm{Z}}_{t+h})'\bm{\Sigma}_h^{-1}(\hat{\bm{Z}}^*_{t+h}-\hat{\bm{Z}}_{t+h})\right],\label{mspe}
\end{align}
where $\hat{\bm{Z}}^*_{t+h}$ and $\hat{\bm{Z}}_{t+h}$ are the $h$-step best linear predictors based on the true and estimated models, respectively, and $\bm{\Sigma}_h$ is the theoretical $h$-step prediction variance with the true model. The PMSE compares forecast performance relative to forecasting with the true parameters, thus has the advantage of resistance to unusual data at the forecasting time points.
The PMSE for only the inner grid points can be obtained by dividing by 25 instead of 49, and using the sub-vector of $\hat{\bm{Z}}^{*(i)}_{t+h}$ and $\hat{\bm{Z}}^{(i)}_{t+h}$, and sub-matrix $\bm{\Sigma}_h$ corresponding to the inner grid points in \eqref{mspe}.

Table~\ref{forecast} reports the empirical PMSE based on the four estimation methods for all 49 locations as well as the inner 25 points. We see that forecasts with $\mathbf{A}$ estimated using the adaptive fused Lasso always have the smallest PMSE.


\section{Application to Daily Wind Speed Data}\label{sec:app}
\begin{figure}[hp!] 
\centering
\includegraphics[width=\textwidth]{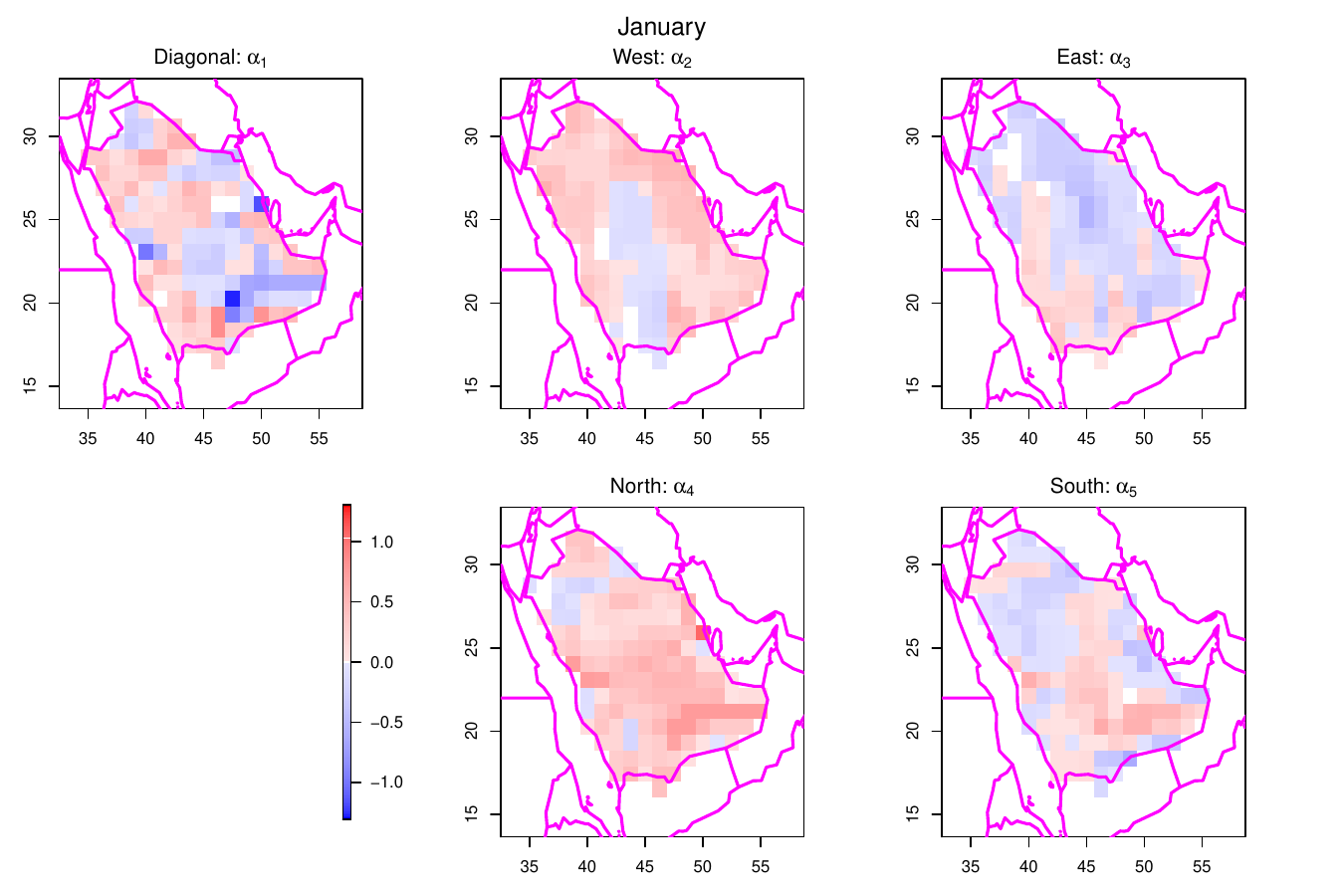}	
\includegraphics[width=\textwidth]{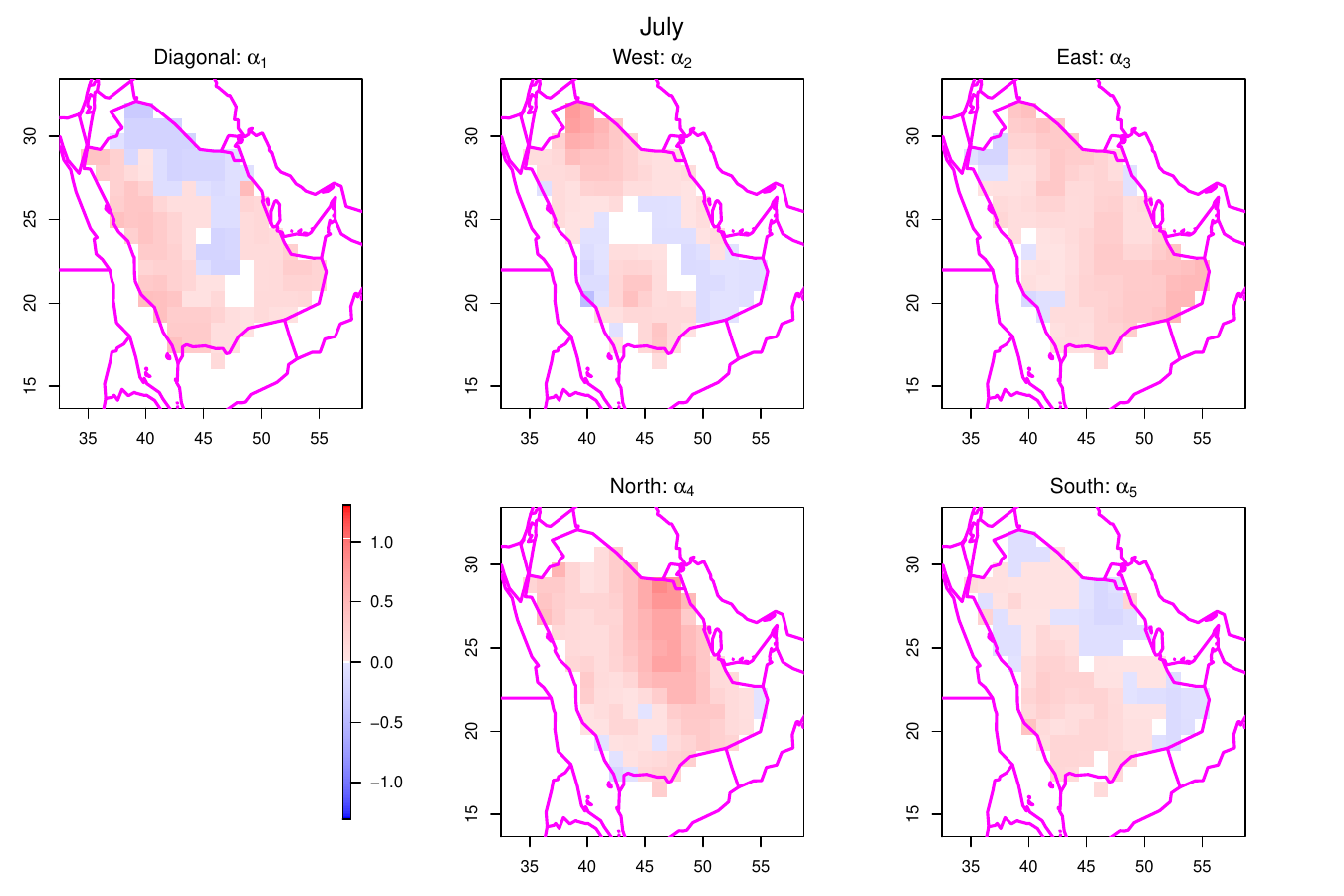}\vspace{-3mm}			
\caption{Maps of the estimated autoregressive coefficients using our algorithm (adaptive fused Lasso) with $K=5$ for the Tukey $g$-and-$h$ transformed residual daily wind speeds on January and July of 86 years (1920-2005) on a grid covering Saudi Arabia.}
\label{fig:wind_est}
\end{figure}

We illustrate the usefulness of our model by a real wind speed data example. The Large Ensemble Project (LENS) dataset is publicly available and consists of 30 ensembles of daily wind speeds over the globe at a spatial resolution of $1.25^\circ$ longitude and $0.94^\circ$ latitude from the year 1920 to 2100 \citep{Kay15} generated from a climate model. We randomly select one ensemble from the dataset and use the historical 86 years (1920-2005) of data at $n=195$ points that covering the domain of Saudi Arabia, plus one additional outer layer of the domain for boundary conditions.
First, we estimate the seasonality of the daily wind speeds at each location by taking the average wind speed across each day of all 86 years at each location. We remove the seasonal effect from each time point and get the residual wind speed for each location. Since the residual daily wind speed is right-skewed, we use the Tukey $g$-and-$h$ transformation, as in \citet{yan18}, to Gaussianize the residuals. We estimate the transformation at each grid point and use the transformed residuals to fit our spatially structured VAR(1) model. Maps of the estimated parameters related to the transformation can be found in \citet{yan18}. 
We use the Gaussianized residual daily wind speeds of January and July in the 86 years ($T=(31-1) \times 86-3=2577$, ignoring points with year jumps and the last 3 time points for forecasting) to fit our model with $K=5$, and use our proposed adaptive fused Lasso algorithm to estimate the coefficients. For this grid configuration, we have $n_I=149, n_B=46$, and hence $m=791$ for the length of $\bm{\alpha}$. Here $\mathbf{\Psi}$ is assumed to be a diagonal matrix and estimated accordingly for better forecasting performance.

Figure~\ref{fig:wind_est} shows the estimation results for the $n_I=149$ inner grid locations.
Comparing maps of these adaptive fused Lasso estimators for January and July, we see distinct patterns of the coefficients. While our method is designed to identify spatially clustered patterns, in this example, it captures mainly smoothly varying coefficients. 
Overall, the autoregressive coefficients in July have larger magnitudes and are smoother than those in January, which indicate stronger temporal dependencies.
The difference can be seen eminently by comparing the coefficients for the northern neighbor between January and July. The stronger dependencies by the northern neighbors in July might be explained by the intensified northerly winds during the monsoon circulation in summer. 
In January, coefficients for the eastern and southern neighbors are negative in the northern part of Saudi Arabia, while in July the patterns are more homogeneous.
We also notice that coefficients for coastal locations are not clustered together and exhibit complex patterns. 
These findings are insightful for explaining the spatiotemporal mechanisms of the wind speed data.
With the estimated parameters, the model can be then used for building stochastic weather generators as an approximation of the computationally expensive climate model. 

Although forecasting is not the main concern for this dataset, for completeness we present the forecast performance by different estimation methods in Table~\ref{data}. Our adaptive fused Lasso outperforms the persistent predictor and the unconstrained VAR(1) estimator. Although occasionally slightly worse than the restricted OLS, the adaptive fused Lasso is more illuminating when visualizing the coefficients. 

\begin{table}[b!]
\centering
\caption{Mean squared prediction error of $h$-step ahead forecasts with $h=1,2,3$ for the 149 inner locations by different estimation methods in January and July.}	
\begin{tabular}{c|c|c|c|c|c}
	& $h$ & Persistent & Unconstrained VAR(1) & Restricted OLS  & Adaptive fused Lasso\\ \hline\hline
	\multirow{3}{*}{Jan}
	& 1 & 1.1510 & 0.7608 & \textbf{0.6951} & 0.6971\\\cline{2-6}
	& 2  & \textbf{1.3327} & 1.6851 & 1.3483 & 1.3369\\\cline{2-6}
	& 3 & 1.5031 & 1.4495 & 1.3039 & \textbf{1.2921}\\\hline\hline
	\multirow{3}{*}{July}
	& 1 & 0.7998 & 0.4852 & 0.3839 & \textbf{0.3838}\\\cline{2-6}
	& 2 & 1.5367 & 1.0933 & 0.4934 & \textbf{0.4932} \\\cline{2-6}
	& 3 & 2.8240 & 1.8587 & 1.0515 & \textbf{1.0432} \\\hline
\end{tabular}
\label{data}
\end{table}

\section{Discussion}\label{sec:diss}
In this paper, we introduced VAR models for stationary time series on gridded locations with sparse and spatially coherent autoregressive coefficients to achieve dimension reduction. Our model is interpretable, flexible, and captures the spatial non-stationarity and essential dynamics of spatiotemporal processes. We detailed our model for the VAR(1) case, developed an adaptive fused Lasso estimation algorithm and derived its asymptotic properties. 
By a simulation study, we showed the satisfying performance of our estimation algorithm as well as the improvement in forecasting gained by the stable estimation.
As illustrated by the data example, our model can be useful to find patterns in the estimated coefficients and thus shed light on the driving mechanisms of the phenomenon of interest over space and time. 
Furthermore, our model designed for gridded spatial data is particularly suitable for building stochastic weather generators. 

For the boundary points, we assumed a simple AR(1) model in order to provide a dynamic mechanism on the boundary for forecasting. This assumption might seem over-simplified, however, estimation of the dynamics for the inner grid points should not be affected much by this assumption as it is done conditional on the boundary variables. If several steps ahead forecast is not of interest, then there is no need to specify a model for the boundary points and the likelihood in \eqref{eq:form} can be replaced by the likelihood conditioning on the boundary points. In this case the transition matrix is of size $n_I \times n$. 
Overall, the effect of edge sites specification is a thorny problem in spatial statistics and deserves further investigation.

Our model is suitable for data rich in space with $n$ up to hundreds of locations, which is typically considered as high-dimensional for a VAR model. For the data example with $n=195$, our estimation algorithm using the `genlasso' package took about 1 hour on a Dell desktop with a 3.30GHz Intel CPU. 
However, with higher dimension, the estimation will take much longer time and become unstable. 
\citet{Zhao20} proposed an algorithm to transform any generalized Lasso problem into a standard Lasso problem, which can be solved using any existing efficient Lasso solver. We implemented their algorithm and found that although faster for large $n$, the solution path is very unstable for $\lambda$ small.
Therefore, a more efficient and stable algorithm solving the generalized Lasso problem is much welcome to be developed and implemented.

Our model can be extended to various directions. Currently, we only apply the adaptive fused Lasso penalty to coefficients corresponding to each of the $K$ neighbors separately in the penalized likelihood (\ref{eq:Phi}). We could further add an adaptive Lasso penalty across the $K$ neighbors, e.g., coefficients corresponding to the left neighbor is expected to be similar to those corresponding to the bottom left and upper left neighbor.
For the non-Gaussian spatiotemporal wind speed data, we estimated the transformation at each location separately. This method could be extended to estimate the parameters related to the transformation and $\bm{\alpha}$ for the underlying spatially structured Gaussian VAR process simultaneously; in this situation, a fused Lasso penalty would also be imposed on the transformation parameters.
It is also possible to extend our method to irregularly spaced data by putting the data on a fine grid with missing observations. For uncertainty quantification, our simulation study and the functional boxplots can be seen as empirical measures of the uncertainty in the generalized Lasso estimates. Recently, \citet{Zhao20} proposed the residual bootstrap method to construct confidence intervals for the generalized Lasso estimates.
At last, observations with measurement error are rarely dealt with for regularized VAR models and could be an interesting topic for further research.

\appendix
\section{Appendix}

\noindent\textbf{Derivation of \eqref{eq:form}}:
$$l(\bm{\alpha})=-\frac{1}{2}\sum_{t=2}^T \bm{Z}_{t-1}'\mathbf{A}'\mathbf{\Psi}^{-1}\mathbf{A}\bm{Z}_{t-1}+\sum_{t=2}^T \bm{Z}'_t\mathbf{\Psi}^{-1}\mathbf{A}\bm{Z}_{t-1}+\mathrm{constant}.$$
For the two terms involved in the above equation,
\begin{align*}
\sum_{t=2}^T \bm{Z}'_t\mathbf{\Psi}^{-1}\mathbf{A}\bm{Z}_{t-1}
=&~ \mathrm{vec}\left(\mathbf{\Psi}^{-1}\sum_{t=2}^{T}\bm{Z}_{t}\bm{Z}_{t-1}'\right)'\mathrm{vec}(\mathbf{A})=\mathrm{vec}\left(\mathbf{\Psi}^{-1}\sum_{t=2}^{T}\bm{Z}_{t}\bm{Z}_{t-1}'\right)'\mathbf{P}\bm{\alpha}, \text{and}\\
\sum_{t=2}^T \bm{Z}_{t-1}'\mathbf{A}'\mathbf{\Psi}^{-1}\mathbf{A}\bm{Z}_{t-1}
=&~ \mathrm{vec}(\mathbf{A})'\left\{\left(\sum_{t=1}^{T-1}\bm{Z}_{t}\bm{Z}_{t}'\right)\otimes\mathbf{\Psi}^{-1}\right\}\mathrm{vec}(\mathbf{A})=\bm{\alpha}'\mathbf{P}'\left\{\left(\sum_{t=1}^{T-1}\bm{Z}_{t}\bm{Z}_{t}'\right)\otimes\mathbf{\Psi}^{-1}\right\}\mathbf{P}\bm{\alpha}.
\end{align*}

Then, it is easy to see that the log-likelihood is a quadratic form of $\bm{\alpha}$, i.e.: $$l(\bm{\alpha})=-\frac{1}{2}\big\|\bm{y}-\bm{X}\bm{\alpha}\big\|_2^2+\mathrm{constant},$$ 
where $\bm{X}$ and $\bm{y}$ satisfy (\ref{eq:x}) and (\ref{eq:y}).  \qed\\

\noindent\textbf{Asymptotic Properties}:

Some notations are needed to state the asymptotic properties. We consider an undirected graph $\mathcal{G}=(\mathcal{V},\mathcal{E})$, where $\mathcal{V}=\{1,\dots,m\}$ is a set of vertices corresponding to the $m=K n_I + n_B$ parameters in $\bm{\alpha}$ to be estimated, and $\mathcal{E}\subset\mathcal{V}\times\mathcal{V}$ is a set of undirected edges that depends on the structure of the spatial grid. Here $\mathcal{V}$ can be partitioned into $n_B+K$ disjoint components, including $n_B$ isolated nodes (corresponding to the non-regularized coefficients in $\bm{\alpha}_\text{B}$) and $K$ identical disjoint components each with $n_I$ vertices and the same inner grid structure (corresponding to the penalized coefficients in $\bm{\alpha}_k$, for $k=1,\dots,K$).
With these notations, the penalty term in (\ref{eq:Phi}) can be rewritten as $$ \lambda\sum_{k=1}^K\sum_{{i}\sim {j}}w_{k,{i},{j}}|\alpha_k(\bs_{i})-\alpha_k(\bs_{j})|=\lambda\sum_{(i,j)\in\mathcal{E}}w_{{i},{j}}|\alpha_i-\alpha_j|,$$ where $w_{{i},{j}}=|\tilde{\alpha}_i-\tilde{\alpha}_j|^{-\gamma}$.

We let $\bm{\alpha}^0$ be the true parameter vector and define $\bm{\xi}=\bm{y}-\bm{X}\bm{\alpha}^0$. We introduce $\mathcal{A}=\left\{(i,j) \in\mathcal{E}: \alpha^0_i=\alpha^0_j, i,j=1,\dots,n_I \right\}$ and consider the subgraph $\mathcal{G}_{\mathcal{A}}=(\mathcal{V},\mathcal{A})$ of  $\mathcal{G}$. We denote by $m_0$ the number of its connected components, that is the number of distinct values in $\bm{\alpha}^0$ supported by $\mathcal{G}$. We further denote by $\mathcal{V}_1,\dots, \mathcal{V}_{m_0}$ the sets of nodes of each connected components of $\mathcal{G}_{\mathcal{A}}$ and set $l_i=\min{\mathcal{V}_i}$ for $i=1,\dots,m_0$. We define $\bm{\alpha}_{\mathcal{A}}^0=(\alpha^0_{l_1},\dots,\alpha^0_{l_{m_0}})'$ and $\bm{X}_\mathcal{A}$ a matrix whose $i$-th column, for $i=1,\dots,m_0$, is $\bm{X}_{\mathcal{A}_i}=\sum_{j\in \mathcal{V}_i}\bm{X}_j$, where $\bm{X}_j$ is the $j$-th column of $\bm{X}$.
We assume that $\bm{X}'\bm{X}/(T-1)\rightarrow\bf{C}$ as $T\rightarrow\infty$ for some positive-definite $m\times m$ matrix $\bf{C}$, which depends on $\bm{\alpha}^0$ and $\mathbf{\Psi}$. We denote by $\bf{C}_{\mathcal{A}}$ the limiting $m_0\times m_0$ matrix of $\bm{X}_{\mathcal{A}}'\bm{X}_{\mathcal{A}}/(T-1)$ as $T\rightarrow\infty$.

We derive the asymptotic properties of the adaptive Lasso estimator $\hat{\bm{\alpha}}$ from solving \eqref{eq:lasso}. We let $\mathcal{A}_n=\left\{(i,j) \in\mathcal{E}: \hat{\alpha}_i=\hat{\alpha}_j,i,j=1,\dots,n_I \right\}$ and $\hat{\bm{\alpha}}_{\mathcal{A}}=(\hat{\alpha}_{l_1},\dots,\hat{\alpha}_{l_{m_0}})'$.

\begin{thm}
	Suppose that $\lambda/\sqrt{T}\rightarrow 0,\, \lambda T^{(\gamma-1)/2}\rightarrow\infty$, and $\bf{C}=\lim_{T\rightarrow\infty}$  $\bm{X}'\bm{X}/(T-1)$ is positive-definite. Then as $T\rightarrow\infty$, the following are satisfied by the adaptive Lasso estimator $\hat{\bm{\alpha}}$:
	\begin{enumerate}
		\item Consistency in grouping: $\Pr(\mathcal{A}_n=\mathcal{A})\rightarrow1$.
		\item Asymptotic normality: $\sqrt{T}\big(\hat{\bm{\alpha}}_{\mathcal{A}}-\bm{\alpha}^0_{\mathcal{A}}\big)\xrightarrow{d}\mathcal{N}_{m_0}(\bm{0},\bf{C}_{\mathcal{A}}^{-1})$.
	\end{enumerate}
\end{thm}

\begin{proof}
	The following proof is obtained by adapting the Proof of Theorem 2 in \citet{adaLasso06} and Theorem 3 in \citet{adaFuLasso13}.
	
	First, it is known that $\hat{\mathbf{\Psi}}$ in \eqref{eq:psi} is a consistent estimator of $\mathbf{\Psi}$, as $T$ goes to infinity; see Chapter 5 of \citet{MTSbook}.
	We define $V_T(\bu)=F(\bm{\alpha}^0)-F(\bm{\alpha}^0+\bu/\sqrt{T})$, with $F$ defined in \eqref{eq:Phi}. It is obvious that $V_T(\bu)$ is minimized at $\sqrt{T}(\hat{\bm{\alpha}}-\bm{\alpha}^0)$ and 
	\begin{align*}
	V_T(\bu)=&\bu'\left(\frac{1}{2T}\bm{X}'\bm{X}\right) \bu-\frac{\bm{\xi}'\bm{X}}{\sqrt{T}}\bu+\frac{\lambda}{\sqrt{T}}\sum_{(i,j)\in\mathcal{A}} w_{{i},{j}}\ \sqrt{T}\left(\left|\alpha_i^0-\alpha_j^0+\frac{u_i-u_j}{\sqrt{T}}\right|-|\alpha_i^0-\alpha_j^0|\right).
	\end{align*}		
	We have
	$$\lambda\ w_{{i},{j}}\left|\alpha_i^0-\alpha_j^0+\frac{u_i-u_j}{\sqrt{T}}\right|-|\alpha_i^0-\alpha_j^0|\xrightarrow{p}\begin{cases}0, &\text{if }\alpha_i^0\neq\alpha_j^0 \text{ or } (\alpha_i^0=\alpha_j^0 \text{ and } u_i=u_j),  \\\infty, &\text{otherwise}.\end{cases}$$
	We denote $\bm{\bu}_{\mathcal{A}}=(\bu_{l_1},\dots,\bu_{l_{m_0}})'$ and then, with an
	application of the Martingale difference central limit theory to $\bm{\xi}'\bm{X}$, we obtain $$V_T(\bu)\xrightarrow{d}V(\bu)=\begin{cases}\frac{1}{2}\bu_\mathcal{A}'\bf{C}_{\mathcal{A}}\bu_\mathcal{A}-\bu_\mathcal{A}'\bf{W}_\mathcal{A}, &\text{if } u_i=u_j \text{ for } (i,j)\in\mathcal{A},\\\infty, &\text{otherwise},\end{cases}$$
	for $\bu \in \mathbb{R}^m$, where ${\textbf W}_\mathcal{A}\sim\mathcal{N}_m(\bm{0},{\textbf C}_{\mathcal{A}})$; $V$ is convex and has a unique minimum satisfying $u_i=u_j$ for all $(i,j)\in\mathcal{A}$ and $\bu_\mathcal{A}=\bf{C}_{\mathcal{A}}^{-1}\bf{W}_\mathcal{A}$. The asymptotic normality part can be derived as in \citet{adaLasso06} by using the epi-convergence results.
	
	For consistency in grouping, we need to show that for all $(i,j)\not\in\mathcal{A}, \Pr((i,j)\in \mathcal{A}_n^c)\rightarrow 1$ and for all $(i,j) \in\mathcal{A}, \Pr((i,j)\in \mathcal{A}_n^c)\rightarrow0$. The first part is implied by the asymptotic normality result. To prove the second part, 
	we 
	apply the subgradient equations for the optimality condition, for $i=1,\dots,m$:
	$$ \bm{X}_i'(\bm{y}-\bm{X}\hat{\bm{\alpha}})=\lambda\sum_{i:(i,j)\in{\mathcal{E}}}w_{{i},{j}}t_{ij},$$ 
	where $t_{ij}=\sign(\hat{\alpha}_i-\hat{\alpha}_j)$ if $\hat{\alpha}_i\neq \hat{\alpha}_j$ and $t_{ij}$ is some real number in $[-1,1]$ if $\hat{\alpha}_i= \hat{\alpha}_j.$
	We prove by contradiction. Suppose that for $\mathcal{V}_k$ that contains at least two vertices, there exist $i,j \in \mathcal{V}_k$ such that $\hat{\alpha}_i\neq \hat{\alpha}_j$. 
	We define $\displaystyle a^{\min}=\min_{i \in \mathcal{V}_k} \hat{\alpha}_i$ and $\mathcal{V}^{\min}=\left\{i: i \in \mathcal{V}_k \text{ and } \hat{\alpha}_i=a^{\min}\right\}$. 
	Summing up the optimality conditions over the indices in $\mathcal{V}^{\min}$, we get:
	$$\sum_{i \in \mathcal{V}^{\min}}\frac{\bm{X}_i'(\bm{y}-\bm{X}\hat{\bm{\alpha}})}{\sqrt{T}}=\frac{\lambda}{\sqrt{T}}\sum_{i \in \mathcal{V}^{\min}} \sum_{\substack{i: (i,j)\in{\mathcal{E}}\\ \alpha_i^0 \neq \alpha_j^0}}\frac{t_{ij}}{|\tilde{\alpha}_i-\tilde{\alpha}_j|^\gamma}+
	\lambda T^{(\gamma-1)/2} \sum_{i \in \mathcal{V}^{\min}} \sum_{\substack{i: (i,j)\in{\mathcal{E}}\\ \alpha_i^0 = \alpha_j^0\\ \hat{\alpha}_j>a^{\min}}}\frac{t_{ij}}{|\sqrt{T}(\tilde{\alpha}_i-\tilde{\alpha}_j)|^\gamma},$$
	where in the right-hand side, the first sum converges to 0 in probability, while the second sum tends to $-\infty$. However, the left-hand side is $O_{p}(1)$, since it can be decomposed as the sum of two asymptotically normal variables as in \citet{adaLasso06} with an application of the martingale central limit theorem. Therefore $\Pr((i,j)\in \mathcal{A}_n^c)\rightarrow0$.
\end{proof}


\begin{thebibliography}{}
\newcommand{\enquote}[1]{``#1''}
\expandafter\ifx\csname natexlab\endcsname\relax\def\natexlab#1{#1}\fi

\bibitem[Ailliot et~al., 2006]{AiMo06}
Ailliot, P., Monbet, V., and Prevosto, M. (2006).
\newblock An autoregressive model with time-varying coefficients for wind
  fields.
\newblock {\em Environmetrics}, 17(2):107--117.

\bibitem[Arnold and Tibshirani, 2020]{genlassoR}
Arnold, T.~B. and Tibshirani, R.~J. (2020).
\newblock {\em genlasso: Path Algorithm for Generalized Lasso Problems}.
\newblock {R} package version 1.5.

\bibitem[Ba{\'n}bura et~al., 2010]{BVAR10}
Ba{\'n}bura, M., Giannone, D., and Reichlin, L. (2010).
\newblock Large {B}ayesian vector auto regressions.
\newblock {\em Journal of Applied Econometrics}, 25(1):71--92.

\bibitem[Basu and Michailidis, 2015]{basu2015}
Basu, S. and Michailidis, G. (2015).
\newblock Regularized estimation in sparse high-dimensional time series models.
\newblock {\em The Annals of Statistics}, 43(4):1535--1567.

\bibitem[Bergmeir and Benítez, 2012]{Ber12}
Bergmeir, C. and Benítez, J.~M. (2012).
\newblock On the use of cross-validation for time series predictor evaluation.
\newblock {\em Information Sciences}, 191:192--213.

\bibitem[Bessac et~al., 2015]{bes15}
Bessac, J., Ailliot, P., and Monbet, V. (2015).
\newblock Gaussian linear state-space model for wind fields in the
  {N}orth-{E}ast {A}tlantic.
\newblock {\em Environmetrics}, 26(1):29--38.

\bibitem[Cressie and Wikle, 2011]{DSTM}
Cressie, N. and Wikle, C.~K. (2011).
\newblock {\em Statistics for Spatio-Temporal Data}.
\newblock Wiley: Hoboken, NJ.

\bibitem[de~Luna and Genton, 2005]{LM05}
de~Luna, X. and Genton, M.~G. (2005).
\newblock Predictive spatio-temporal models for spatially sparse environmental
  data.
\newblock {\em Statistica Sinica}, 15:547--568.

\bibitem[Fan and Zhang, 1999]{fan1999}
Fan, J. and Zhang, W. (1999).
\newblock Statistical estimation in varying coefficient models.
\newblock {\em Annals of Statistics}, 27(5):1491--1518.

\bibitem[Gelfand et~al., 2003]{SVC03}
Gelfand, A.~E., Kim, H.-J., Sirmans, C.~F., and Banerjee, S. (2003).
\newblock Spatial modeling with spatially varying coefficient processes.
\newblock {\em Journal of the American Statistical Association},
  98(462):387--396.

\bibitem[Genton et~al., 2014]{sufbplot}
Genton, M., Johnson, C., Potter, K., Stenchikov, G., and Sun, Y. (2014).
\newblock Surface boxplots.
\newblock {\em Stat}, 3(1):1--11.

\bibitem[Hsu et~al., 2008]{Hsu08}
Hsu, N.-J., Hung, H.-L., and Chang, Y.-M. (2008).
\newblock Subset selection for vector autoregressive processes using {L}asso.
\newblock {\em Computational Statistics and Data Analysis}, 52(7):3645--3657.

\bibitem[Huang et~al., 2010]{SpL10}
Huang, H.-C., Hsu, N.-J., Theobald, D.~M., and Breidt, F.~J. (2010).
\newblock Spatial {L}asso with applications to {GIS} model selection.
\newblock {\em Journal of Computational and Graphical Statistics},
  19(4):963--983.

\bibitem[Kastner and Huber, 2020]{BVAR20}
Kastner, G. and Huber, F. (2020).
\newblock Sparse {B}ayesian vector autoregressions in huge dimensions.
\newblock {\em Journal of Forecasting}, 39:1142--1165.

\bibitem[Katzfuss and Cressie, 2012]{kat12}
Katzfuss, M. and Cressie, N. (2012).
\newblock Bayesian hierarchical spatio-temporal smoothing for very large
  datasets.
\newblock {\em Environmetrics}, 23(1):94--107.

\bibitem[Kay et~al., 2015]{Kay15}
Kay, J.~E., Deser, C., Phillips, A., Mai, A., Hannay, C., Strand, G.,
  Arblaster, J.~M., Bates, S.~C., Danabasoglu, G., Edwards, J., Holland, M.,
  Kushner, P., Lamarque, J.-F., Lawrence, D., Lindsay, K., Middleton, A.,
  Munoz, E., Neale, R., Oleson, K., Polvani, L., and Vertenstein, M. (2015).
\newblock The community earth system model {(CESM)} large ensemble project: A
  community resource for studying climate change in the presence of internal
  climate variability.
\newblock {\em Bulletin of the American Meteorological Society},
  96(8):1333--1349.

\bibitem[Ke et~al., 2015]{KeFan15}
Ke, Z.~T., Fan, J., and Wu, Y. (2015).
\newblock Homogeneity pursuit.
\newblock {\em Journal of the American Statistical Association},
  110(509):175--194.

\bibitem[Korobilis and Pettenuzzo, 2019]{BVAR19}
Korobilis, D. and Pettenuzzo, D. (2019).
\newblock Adaptive hierarchical priors for high-dimensional vector
  autoregressions.
\newblock {\em Journal of Econometrics}, 212(1):241--271.

\bibitem[Li and Sang, 2019]{LiSang19}
Li, F. and Sang, H. (2019).
\newblock Spatial homogeneity pursuit of regression coefficients for large
  datasets.
\newblock {\em Journal of the American Statistical Association},
  114(527):1050--1062.

\bibitem[L{\"u}tkepohl, 2005]{MTSbook}
L{\"u}tkepohl, H. (2005).
\newblock {\em New Introduction to Multiple Time Series Analysis}.
\newblock Springer-Verlag: Berlin.

\bibitem[Monbet and Ailliot, 2017]{MoAi17}
Monbet, V. and Ailliot, P. (2017).
\newblock Sparse vector {M}arkov switching autoregressive models. {A}pplication
  to multivariate time series of temperature.
\newblock {\em Computational Statistics and Data Analysis}, 108:40--51.

\bibitem[Mu et~al., 2018]{SVC18}
Mu, J., Wang, G., and Wang, L. (2018).
\newblock Estimation and inference in spatially varying coefficient models.
\newblock {\em Environmetrics}, 29(1):e2485.

\bibitem[Ngueyep and Serban, 2015]{Ngue15}
Ngueyep, R. and Serban, N. (2015).
\newblock Large-vector autoregression for multilayer spatially correlated time
  series.
\newblock {\em Technometrics}, 57(2):207--216.

\bibitem[{R Development Core Team}, 2020]{R19}
{R Development Core Team} (2020).
\newblock {\em R: A Language and Environment for Statistical Computing}.
\newblock R Foundation for Statistical Computing, Vienna, Austria.

\bibitem[Rao, 2008]{Rao08}
Rao, S.~S. (2008).
\newblock Statistical analysis of a spatio-temporal model with
  location-dependent parameters and a test for spatial stationarity.
\newblock {\em Journal of Time Series Analysis}, 29(4):673--694.

\bibitem[Schweinberger et~al., 2017]{Add17}
Schweinberger, M., Babkin, S., and Ensor, K.~B. (2017).
\newblock High-dimensional multivariate time series with additional structure.
\newblock {\em Journal of Computational and Graphical Statistics},
  26(3):610--622.

\bibitem[Shen and Huang, 2010]{GroupL10}
Shen, X. and Huang, H.-C. (2010).
\newblock Grouping pursuit through a regularization solution surface.
\newblock {\em Journal of the American Statistical Association},
  105(490):727--739.

\bibitem[Sun and Genton, 2011]{SG11}
Sun, Y. and Genton, M.~G. (2011).
\newblock Functional boxplots.
\newblock {\em Journal of Computational and Graphical Statistics},
  20(2):316--334.

\bibitem[Sun et~al., 2016]{Sun16}
Sun, Y., Wang, H.~J., and Fuentes, M. (2016).
\newblock Fused adaptive {L}asso for spatial and temporal quantile function
  estimation.
\newblock {\em Technometrics}, 58(1):127--137.

\bibitem[Tagle et~al., 2019]{Tagle17}
Tagle, F., Castruccio, S., Crippa, P., and Genton, M.~G. (2019).
\newblock A non-{G}aussian spatio-temporal model for daily wind speeds based on
  a multivariate skew-$t$ distribution.
\newblock {\em Journal of Time Series Analysis}, 40:312--326.

\bibitem[Tibshirani et~al., 2005]{fused05}
Tibshirani, R., Saunders, M., Rosset, S., Zhu, J., and Knight, K. (2005).
\newblock Sparsity and smoothness via the fused lasso.
\newblock {\em Journal of the Royal Statistical Society: Series B},
  67(1):91--108.

\bibitem[Tibshirani and Taylor, 2011]{genlasso11}
Tibshirani, R.~J. and Taylor, J. (2011).
\newblock The solution path of the generalized lasso.
\newblock {\em The Annals of Statistics}, 39(3):1335--1371.

\bibitem[Viallon et~al., 2013]{adaFuLasso13}
Viallon, V., Lambert-Lacroix, S., H{\"o}fling, H., and Picard, F. (2013).
\newblock Adaptive generalized fused-{L}asso: Asymptotic properties and
  applications.
\newblock {\em HAL preprint}, hal-00813281.

\bibitem[Wikle et~al., 1998]{Wikle1998}
Wikle, C.~K., Berliner, L.~M., and Cressie, N. (1998).
\newblock Hierarchical {B}ayesian space-time models.
\newblock {\em Environmental and Ecological Statistics}, 5(2):117--154.

\bibitem[Wikle et~al., 2001]{Wi01}
Wikle, C.~K., Milliff, R.~F., Nychka, D., and Berliner, L.~M. (2001).
\newblock Spatiotemporal hierarchical bayesian modeling: Tropical ocean surface
  winds.
\newblock {\em Journal of the American Statistical Association},
  96(454):382--397.

\bibitem[Yan and Genton, 2019]{yan18}
Yan, Y. and Genton, M.~G. (2019).
\newblock Non-{G}aussian autoregressive processes with {T}ukey $g$-and-$h$
  transformations.
\newblock {\em Environmetrics}, 30:e2503.

\bibitem[Zhao and Bondell, 2020]{Zhao20}
Zhao, Y. and Bondell, H. (2020).
\newblock Solution paths for the generalized lasso with applications to
  spatially varying coefficients regression.
\newblock {\em Computational Statistics and Data Analysis}, 142:106821.

\bibitem[Zou, 2006]{adaLasso06}
Zou, H. (2006).
\newblock The adaptive lasso and its oracle properties.
\newblock {\em Journal of the American Statistical Association},
  101(476):1418--1429.

\end{thebibliography}
\end{document}